\RequirePackage{fix-cm}
\documentclass[twocolumn,epjc3]{svjour3}
\smartqed  
\RequirePackage{graphicx}

\usepackage{amssymb,graphics}
\usepackage{graphics,epsfig,subfigure}
\usepackage{epstopdf}
\usepackage{graphicx}
\usepackage{dcolumn,color}
\usepackage{bm}
\usepackage{epsfig}
\usepackage{color,framed}

\journalname{Eur. Phys. J. C}
\begin{document}
\renewcommand{\baselinestretch}{1.3}
\newcommand\beq{\begin{equation}}
\newcommand\eeq{\end{equation}}
\newcommand\beqn{\begin{eqnarray}}
\newcommand\eeqn{\end{eqnarray}}
\newcommand\fc{\frac}
\newcommand\lt{\left}
\newcommand\rt{\right}
\newcommand\pt{\partial}
\title{The structure of $f(R)$-brane model}



\author{Zeng-Guang Xu\thanksref{e1,addr1}
        \and Yuan Zhong\thanksref{e2,addr1,addr2}
         \and Hao Yu\thanksref{e3,addr1}
         \and Yu-Xiao Liu\thanksref{e4,addr1,addr3}
        }

\thankstext{e1}{e-mail:xuzg12@lzu.edu.cn}
\thankstext{e2}{e-mail:zhongy2009@lzu.edu.cn}
\thankstext{e3}{e-mail:yuh13@lzu.edu.cn}
\thankstext{e4}{e-mail:liuyx@lzu.edu.cn, corresponding author}


\institute{Institute of Theoretical Physics,
            Lanzhou University, Lanzhou 730000,
            People's Republic of China \label{addr1}
           \and
           IFAE, Universitat Aut$\grave{\textrm{o}}$noma de Barcelona, 08193 Bellaterra, Barcelona, Spain \label{addr2}
           \and
           Key Laboratory for Magnetism and Magnetic Materials of the MoE,
                 Lanzhou University, Lanzhou 730000, China \label{addr3}
}

\date{Received: date / Accepted: date}

\maketitle

\begin{abstract}
Recently, a family of interesting analytical brane solutions were found in $f(R)$ gravity with $f(R)=R+\alpha R^2$ in Ref. [Phys. Lett. B 729, 127 (2014)]. In these solutions, inner brane structure can be turned on by tuning the value of the parameter $\alpha$.
In this paper, we investigate how the parameter $\alpha$ affects the localization and the quasilocalization of the tensorial gravitons around these solutions. It is found that, in a range of $\alpha$, despite the brane has an inner structure, there is no graviton resonance.
However, in some other regions of the parameter space, although the brane has no internal structure, the effective potential for the graviton KK modes has a singular structure, and there exists a series of graviton resonant modes. The  contribution of the  massive graviton KK modes to the Newton's law of gravity is discussed shortly.
\end{abstract}

\section{Introduction}
The braneworld scenarios~\cite{Antoniadis1990,AntoniadisArkani-HamedDimopoulosDvali1998,Arkani-HamedDimopoulosDvali1998a,RandallSundrum1999,RandallSundrum1999a}, have attracted more and more attention. Because they present new insights and solutions for many issues such as the hierarchy problem, the cosmological problem, the nature of dark matter and dark energy, and so on. In some braneworld scenarios, all the fields in the standard model are assumed to be trapped on a submanifold (called brane) in a higher-dimensional spacetime (called bulk), and only gravity transmits in the whole bulk.
So a natural and interesting question is how to reproduce the effective four-dimensional Newtonian gravity from a bulk gravity.

In models with compacted extra dimension, such as the ADD braneworld model~\cite{AntoniadisArkani-HamedDimopoulosDvali1998,Arkani-HamedDimopoulosDvali1998a}, the effective gravity on the brane is transmitted by the massless graviton Kaluza-Klein (KK) mode (also called the graviton zero mode). Since the graviton zero mode is separated from the first KK excitation by a mass gap, gravity is effectively four-dimensional at low energy.
While in models with infinitely large extra dimensions, like the one proposed by Randall and Sundrum (the RS2 model)~\cite{RandallSundrum1999a}, although the graviton spectrum is gapless now, the effective gravity on the brane is shown to be the Newtonian gravity plus an subdominant correction, so the low energy effective gravity is also four-dimensional ~\cite{Lykken1999nb,Dvali:2000rv,Csaki:2000fc,Csaki:2000pp}. This is because the wave functions of all the massive graviton KK modes are suppressed on the brane, so, the massive modes only contribute a small correction to the Newtonian gravity at large distance.

Another class of interesting models with infinite extra dimensions is the so called thick branes. In these modes, the original singular thin brane in the RS2 model is replaced by some smooth thick domain walls generated by one or a few background scalars~\cite{Gremm2000,KehagiasTamvakis2001,CsakiErlichHollowoodShirman2000}.
One of the interesting features of thick domain wall brane is that the massive graviton modes feel an effective volcanolike potential, which might support some resonances. This feature was first noticed in Ref.~\cite{Gremm2000}.

Such resonant modes of graviton can be interesting both phenomenlogically and theoretically.
Phenomenlogically, the wave function of a graviton resonance peaks at the location of brane and behaviors as a plain wave in the infinity of the extra dimension. As compared to the RS2 model, the massive modes of a thick brane might contribute a different correction to the Newtonian gravity. On the theoretical aspect, metastable massive graviton in thick brane models provides an alternative for massive gravity theory~\cite{Hinterbichler2012,Rham2014}. Unfortunately, the early proposal for thick brane~\cite{Gremm2000,KehagiasTamvakis2001,CsakiErlichHollowoodShirman2000} failed in finding massive graviton resonance.

To construct a model that support graviton resonance, one has to tune the shape of the effective potential for the graviton. For typical thick brane models, where the gravity is taken as general relativity, the only possible way is to tune the shape of the warp factor. Some successful models can be found in Refs.~\cite{GuoLiuZhaoChen2012,XieYangZhao2013,CruzSousaMalufAlmeida2014,ZhongLiuZhao2014}.
There is another way, however, to tune the effective potential for the graviton if the gravity is described by a more general theory. For example, in $f(R)$ gravity (see~\cite{SotiriouFaraoni2010,DeTsujikawa2010,NojiriOdintsov2011} for comprehensive reviews on $f(R)$ gravity and its applications in cosmology), the effective potential is determined by both the warp factor and the form of $f(R)$ (see Ref.~\cite{ZhongLiuYang2011} for details). So, in principle, graviton resonances can be turned on by tuning gravity.

However, the construction of a thick $f(R)$-brane model is not easy in practice (see Refs.~\cite{ParryPichlerDeeg2005,AfonsoBazeiaMenezesPetrov2007,DeruelleSasakiSendouda2008,BalcerzakDabrowski2008,DzhunushalievFolomeevKleihausKunz2010,LiuZhongZhaoLi2011,HoffDias2011,LiuLuWang2012,CaramesGuimaraesSilva2012,BazeiaMenezesPetrovSilva2013,BazeiaLobaoMenezesPetrovSilva2014} for works on $f(R)$-branes). First of all, the dynamical equation in $f(R)$-gravity is of fourth order. So, traditional methods that help us to find analytical brane solutions in second order systems, such as the superpotential method (also known as the first order formalism~\cite{DeWolfeFreedmanGubserKarch2000,AfonsoBazeiaLosano2006}) do not work in a general $f(R)$-brane model. Usually one can only solve the system either by using numerical method~\cite{DzhunushalievFolomeevKleihausKunz2010}, or by imposing strict constraints on the model, for example, by assuming the scalar curvature to be a constant~\cite{AfonsoBazeiaMenezesPetrov2007}.

The linearization of a $f(R)$-brane model is also a challenging task. Naively, the linear perturbations around an arbitrary $f(R)$-brane solution should also satisfy some fourth order differential equations. However, in Ref.~\cite{ZhongLiuYang2011}, the authors found that the tensor perturbation perturbation equation can be finally rewritten as a second order Schr\"odinger-like equation. The results of Ref.~\cite{ZhongLiuYang2011} enable us to analyse the graviton mass spectrum of any thick $f(R)$-branes solutions. For example, in Ref.~\cite{LiuZhongZhaoLi2011}, the authors constructed an analytical thick brane solution in a model with $f(R)=R+\alpha R^2$. The solution is stable against the tensor perturbation. The localization of zero modes of both graviton and fermion is also shown to be possible. Unfortunately, no graviton resonance was found in the model of Ref.~\cite{LiuZhongZhaoLi2011}. Recently, a more general analytical thick $f(R)$-brane solution was reported in Ref.~\cite{BazeiaLobaoMenezesPetrovSilva2014}. The solution of Ref.~\cite{BazeiaLobaoMenezesPetrovSilva2014} is a generalization of the one in~\cite{LiuZhongZhaoLi2011}. The model in Ref.~\cite{BazeiaLobaoMenezesPetrovSilva2014} has an interesting feature: inner brane structure appears for a particular range of the parameter $\alpha$. Usually, the appearance of inner brane structure is accompanied by graviton resonances~\cite{GuoLiuZhaoChen2012,XieYangZhao2013,CruzSousaMalufAlmeida2014,ZhongLiuZhao2014}. So it is interesting to see if it is possible to find graviton resonances in the model of~\cite{BazeiaLobaoMenezesPetrovSilva2014}, and what is the relation between the inner brane structure and the graviton resonance. These two questions constitute the motivation for the present work.

In the next section~\ref{SecModel}, we briefly review the model and corresponding solutions in Refs. \cite{LiuZhongZhaoLi2011} and \cite{BazeiaLobaoMenezesPetrovSilva2014}. Then, in section \ref{main}, we study the localization of the graviton zero mode and the condition for graviton resonances in the model of~\cite{BazeiaLobaoMenezesPetrovSilva2014}. The correction from the the massive graviton KK modes at small distance is discussed in order to compare with the constraints of breaking the Newton's inverse square law from experiments given in Ref.~\cite{Long:2002wn}. The conclusion and discussions will be given in section \ref{Conclusion}.

\section{Review of the $f(R)$-brane model and solutions}
\label{SecModel}

We start with the five-dimensional action of the $f(R)$ gravity minimally coupled with a canonical scalar field
\begin{eqnarray}
  S=\int d^4x dy\sqrt {-g}\left(\frac{1}{2\kappa_5^2}f(R)
  -\frac12\partial^M\phi\partial_M\phi-V(\phi)\right),
  \label{action}
\end{eqnarray}
where $f(R)$ is a function of the scalar curvature $R$, and {$\kappa_5^2=2M_*^3$ with $M_*$ the fundamental five-dimensional Planck mass. In the following, we set $M_*=1$.} The signature of the metric is taken as $(-, +, +, +, +)$, and the bulk coordinates are denoted by Capital Latin indices, $M, N, \cdots=0, 1, 2, 3, 4$, and the brane coordinates are denoted Greek indices, $\mu,\nu, \cdots=0,1,2,3$.

We are interested in the static Minkowski brane embedded in a five-dimensional spacetime, so the line element is assumed as
\begin{eqnarray}
~~~~~~~~~~~~ds^{2}=e^{2A(y)}\eta_{\mu\nu}dx^{\mu}dx^{\nu}+dy^{2}, \label{Metric}
\end{eqnarray}
where $e^{2A(y)}$ is the warp factor, $y=x^{4}$ stands for the extra dimension, and $\eta_{\mu\nu}$ is the induced metric on the brane.

For static brane solutions with the setup (\ref{Metric}), the background scalar field $\phi$ is only the function of the extra dimension, i.e., $\phi=\phi(y)$.
Therefore, the equations of motion for the $f(R)$-brane system are
\begin{eqnarray}
\label{phi}
 \phi''+4A'\phi' &=&V_{\phi},\\
\label{EE1}
 f\!+\!2f_R\left(4A'^2 \!+\! A''\right)
  \!-\! 6f'_RA'\!-\!2f''_R&=&\kappa_5^2(\phi'^2\!+\!2V),\\
\label{EE2}
 -8f_R\left(A''+A'^2\right)+8f'_RA'
  -f&=&\kappa_5^2(\phi'^2\!-\!2V),
\end{eqnarray}
where the prime denotes the derivative with respect to $y$, $f_{R}{\equiv}df(R)/dR$, and $V_{\phi}{\equiv}dV(\phi)/d\phi$.

In Ref. \cite{LiuZhongZhaoLi2011}, a toy model with
\begin{eqnarray}
~~~~~~~~~~~~~~~~~~~~~f(R)=R+\alpha R^2  \label{FR}
\end{eqnarray}
and the $\phi^4$ potential
\begin{eqnarray}
~~~~~~~~~~~~~~~V(\phi)=\lambda(\phi^2-v^2)^2+\Lambda_5 \label{Vphi4}
\end{eqnarray}
was considered, where $\lambda>0$ is the self coupling constant of the scalar field, and $v$ is the vacuum expectation value of the scalar field. An analytical solution was found in Ref. \cite{LiuZhongZhaoLi2011}
\begin{eqnarray}
~~~~~~~~~~~~~~~~~~~~~\textrm{e}^{A(y)}&=&\textrm{sech}(ky), \label{Ay1}\\
 \phi (y)&=&v \tanh(k y),\label{phiy1}
\end{eqnarray}
where the parameters are given by
\begin{eqnarray}
 ~~~~~~~~~~~~k&=&\sqrt{\frac{3}{232 \alpha }},~~ \quad
 \lambda =\frac3{784} \frac{\kappa
_5^2}{\alpha},\\
 v &=&  7 \sqrt{\frac{3}{29\kappa _5^2}},\quad
\Lambda_5=-\frac{477}{6728 }\frac{1} {\alpha\kappa _5^2}. \label{Lambda5}
\end{eqnarray}
The scalar field satisfies $\phi(0)=0$ and $\phi(\pm\infty)=\pm v$, and the potential reaches the minimum (the vacuum) at $\phi=\pm v$.
The energy density $\rho=T_{00}=e^{2A}(\frac12\phi'^2+V(\phi))$ peaks at the location of the brane, $y=0$, and trends to vanish at the boundary of the extra dimension. The brane is embedded in an anti-de Sitter spacetime.

The tensor perturbation of the above brane solution has been analyzed in Ref. \cite{LiuZhongZhaoLi2011}. It was shown that the solution is stable against the tensor perturbation and the gravity zero mode is localized on the brane. Furthermore, it was found that although there are fermion resonant KK modes on the brane, there are no graviton resonant modes \cite{LiuZhongZhaoLi2011}.

\begin{figure}
\begin{center}
\subfigure[$B=1$]{\label{EnergyDensityB_1}
\includegraphics[width=0.35\textwidth]{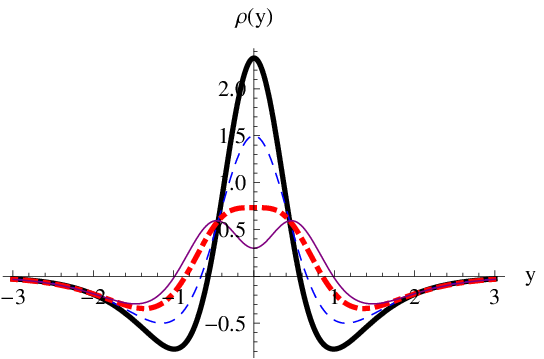}}
\subfigure[$B=4$]{\label{EnergyDensityB_4}
\includegraphics[width=0.35\textwidth]{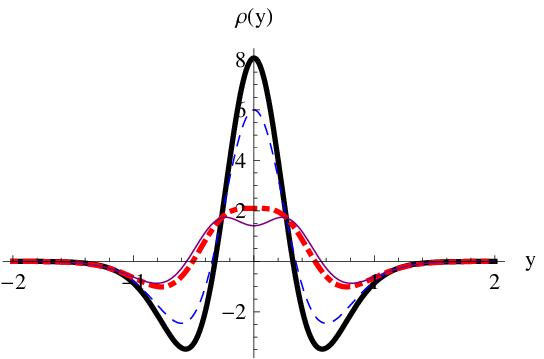}}
\end{center}
\caption{The energy density $\rho(y)$ of the brane system. The parameters are set to $B=1$ (a) and $B=4$ (b), $k=1$, and $\alpha=\alpha_{1}$ (solid thin purple line), $\alpha=\alpha_{s}$ (dotted-dashing red line), $\alpha=0$ ( dashed blue line), and $\alpha_{2}$ (thick black line). } \label{figEnergyDensity}
\end{figure}

Recently, a general solution was constructed in Ref. \cite{BazeiaLobaoMenezesPetrovSilva2014} with the same $f(R)$ given in Eq. (\ref{FR}). The warp factor is assumed as the general form of (\ref{Ay1}) with two positive parameters $B$ and $k$:
\begin{eqnarray}
 ~~~~~~~~~~~~~~~~~~~ \textrm{e}^{A(y)}=\textrm{sech}^{B}(ky). \label{Ay}
\end{eqnarray}
The corresponding Ricci tensor at the boundary of the extra dimension is given by
\begin{eqnarray}
 ~~~~ R_{MN}(y\rightarrow\pm\infty)\rightarrow -4B^2 k^2  g_{MN}\equiv{\Lambda_{\textrm{eff}}}~g_{MN},
\end{eqnarray}
from which we can see that the background spacetime is asymptotical AdS with the effective cosmological constant ${\Lambda_{\textrm{eff}}}=-4B^2 k^2$. The curvature is
\begin{eqnarray}
 ~~~~ R(y)=-4Bk^{2}[5B-5(B+2)\textrm{sech}^{2}(ky)].
\end{eqnarray}
Then from Eq.~(\ref{EE1}) we can give the derivative of the scalar field:
\begin{eqnarray}
\label{phi}
{\phi'^{2}}&=&{Bk^2}\textrm{sech}^{2}(ky)\Big\{\frac{3}{2}-4\alpha k^{2}\Big[5B^{2}+16B+8 \nonumber \\
  &&-(5B^{2}+32B+12)\textrm{sech}^{2}(ky)\Big] \Big\}.
\end{eqnarray}
Note that the $\alpha$ here corresponds to $-\alpha$ in Ref.~\cite{BazeiaLobaoMenezesPetrovSilva2014}. Thus, $\phi'^{2}\geq0$ implies
\begin{eqnarray}
\alpha_{1}\!\equiv-\frac{3}{32(1\!+\!4B)k^2}\!\leq\!\alpha \!\leq\!\frac{3}{8(8\!+\!16B\!+\!5B^2)k^2} \equiv \alpha_{2}.
\label{ConstrainOnalpha}
\end{eqnarray}
The energy-momentum tensor density is
\begin{eqnarray}\label{rho}
\rho \!\!&=&\!\! e^{2A(y)}\left(\frac{1}{2}\partial^{M}\phi\partial_{M}\phi+V(\phi)\right)\nonumber\\
\!\!&=&\!\! Bk^{2} \textrm{sech}^{2B}(ky)
   \Big\{-3B + 3\Big(B+\frac{1}{2}\Big)\textrm{sech}^{2}(ky)\nonumber\\
&&\!\! +4\alpha k^{2}
   \Big[5B^{3} -(10B^{3}+37B^{2}+32B+8)\textrm{sech}^{2}(ky)\nonumber\\
&&\!\!     +(5B^{3}+37B^{2}+44B+12)\textrm{sech}^{4}(ky)\Big]\Big\}.
\end{eqnarray}
Solving $\frac{d^2\rho}{dy^2}\big|_{y=0}=0$ results in
\begin{eqnarray}
 ~~~~ ~~~~ \alpha=\alpha_{s}\equiv-\frac{3+9B}{8k^{2}(16+60B+49B^{2})},
\end{eqnarray}
where $\alpha_{1}<\alpha_{s}<\alpha_{2}$ for finite $B$. So $y=0$ is an inflection point of $\rho$ when $\alpha=\alpha_{s}$, and the brane will have an internal structure when $\alpha\leq\alpha_{s}$. The energy density $\rho(y)$ of the brane system is shown in Fig.~\ref{figEnergyDensity}.

For an arbitral $B$ and a fixed $\alpha$, the analytical solutions for the scalar field and scalar potential were found in Ref.~\cite{BazeiaLobaoMenezesPetrovSilva2014}:
when $\alpha=\alpha_{1}$, the solution is
\begin{eqnarray}
~~~~~~~~~\phi(y)\!\!&=&\!\! v_1 [1-\textrm{sech}(k y)]\textrm{sign}(y),\\
V(\phi)\!\!&=&\!\!c_2 \big(|\phi|-v_1 \big)^{2}
    \left[\big(|\phi|-v_1 \big)^{2}-c_1 \right]-c_0;
\end{eqnarray}
when $\alpha=0$, we have
\begin{eqnarray}
\phi(y)\!\!&=&\!\!\sqrt{6B}\arctan\left[\tanh\left(\frac{k y}{2}\right)\right],\\
V(\phi)\!\!&=&\!\!-3B^{2}k^{2}+3B k^{2}\left(B+\frac{1}{4}\right)\cos^{2}\left(\sqrt{\frac{2}{3B}}\phi\right);~~~
\end{eqnarray}
when $\alpha=\alpha_{2}$, the result is just the one found in Ref. \cite{LiuZhongZhaoLi2011}:
\begin{eqnarray}
~~~~~~~~~~~~\phi(y)\!\!&=&\!\! v_2 \tanh(k y), \label{phiyalpha2}\\
V(\phi)\!\!&=&\!\!\lambda_1(\phi^2-v_2^2)^2-\lambda_0. \label{Vphialpha2}
\end{eqnarray}
Here $c_i$, $v_i$, and $\lambda_i$ are positive parameters determined by $B$ and $k$. Here, we only list the expressions of $v_i$:
\begin{eqnarray}
~~~~~~~~~~~~~~~ v_1 &=& \sqrt{\frac{3B(6+B)(2+5B)}{8+32B}}, \label{v1} \\
 v_2 &=& \sqrt{\frac{3B(6+B)(2+5B)}{16+2B(16+5B)}}. \label{v2}
\end{eqnarray}
Now, it is clear that, when $B=1$ and $\alpha=\alpha_2=\frac{3}{232 k^2}$, the exact solution described by Eqs. (\ref{Ay}), (\ref{phiyalpha2}), (\ref{Vphialpha2}), and (\ref{v2}) is just the one given in Eqs. (\ref{FR})-(\ref{Lambda5}), for which the brane has no internal structure.

\section{Localization and resonant KK modes of the tensor fluctuation}

\label{main}

In this section, we investigate the stability of solution against tensor fluctuation of the metric as well as localization of gravity on the brane. Especially, we will find that the effective potential for the KK modes of the tensor fluctuation has a rich structure and it will support some resonant KK modes when $B$ is large enough and $\alpha>0$.

We start with the transverse-traceless (TT) tensor perturbations of the background spacetime:
\begin{eqnarray}
~~~~~~~~ds^2=\textrm{e}^{2A(y)}(\eta_{\mu\nu}+h_{\mu\nu})dx^\mu dx^\nu+dy^2,
\end{eqnarray}
where $h_{\mu\nu}=h_{\mu\nu}(x^{\rho}, y)$ depends on all the spacetime coordinates and satisfies the TT condition
\begin{eqnarray}
~~~~~~~~~~~~~~~~\eta^{\mu\nu}h_{\mu\nu}=\partial_{\mu}h^{\mu}_{\nu}=0.
\end{eqnarray}
It can be shown that the TT tensor perturbations $h_{\mu\nu}$ is decoupled from the scalar and vector perturbations of the metric as well as the perturbation of the scalar field $\delta\phi=\tilde{\phi}(x^{\rho}, y)$. The perturbed Einstein equations for the TT tensor perturbations are given by \cite{ZhongLiuYang2011}
\begin{eqnarray}
~~~~~~~~~~~~~~~~\square^{(5)}h_{\mu\nu}=\frac{f_R'}{f_R}\partial_y h_{\mu\nu},
\end{eqnarray}
or, equivalently,
\begin{eqnarray}
\label{eqEEFinal}
\left(a^{-2}\square^{(4)}h_{\mu\nu}+4\frac{a'}{a}h_{\mu\nu}'+h_{\mu \nu }''\right)f_R+h_{\mu\nu}'f_R'=0,~~
\end{eqnarray}
where $a(y)=\textrm{e}^{A(y)}$.
By making the coordinate transformation $dz=a^{-1}dy$, Eq. (\ref{eqEEFinal}) becomes
\begin{eqnarray}
~~~~\left[\partial_z^{2}+\left(3\frac{\partial_z a}{a}+\frac{\partial_z f_R}{f_R}\right)\partial_z+\square^{(4)}\right]h_{\mu\nu}=0.  \label{eqEEFinal2}
\end{eqnarray}
Then, by doing the decomposition
\begin{eqnarray}
~~~~~~~~~~h_{\mu\nu}(x^{\rho},z)=(a^{-3/2}f_R^{-1/2})\epsilon_{\mu\nu}(x^{\rho})\psi(z),
\end{eqnarray}
where $\epsilon_{\mu\nu}(x^{\rho})$ satisfies the TT condition $\eta^{\mu\nu}\epsilon_{\mu\nu}=0=\partial_\mu
\epsilon^{~\mu}_\nu$, we can get from (\ref{eqEEFinal2}) the equation for the KK modes $\psi(z)$ of the tensor perturbations, which is a Schr\"odinger-like equation \cite{ZhongLiuYang2011}:
\begin{eqnarray}
\label{Schrodinger}
~~~~~~~~~~\left[-\partial_z^2+W(z)\right]\psi(z)=m^2\psi(z),  \label{SchrodingerEq}
\end{eqnarray}
where the effective potential $W(z)$ is
\begin{eqnarray}
\label{Schrodingerpotential}
~~~~~~~~W(z)\!\!&=&\!\!\frac34\frac{(\partial_{z}a)^2}{a^2}+\frac32\frac{\partial_z^{2}a}{a}+\frac32\frac{\partial_{z}a \partial_{z}f_R}{a f_R}\nonumber \\
  &-&\!\!\frac14\frac{(\partial_z f_R)^2}{f_R^2}+\frac12\frac{\partial_z^{2}f_R}{f_R}.  \label{Wz}
\end{eqnarray}
Equation (\ref{SchrodingerEq}) can be factorized as
\begin{eqnarray}
  ~~~~~~~~~~~~~~~~\mathcal{K}\mathcal{K}^{\dagger} \psi(z) =m^2\psi(z),  \label{SchrodingerEq2}
\end{eqnarray}
with
 \begin{eqnarray}
 ~~~~~~~~~~~~~~~~ \mathcal{K}&=&+\partial_z
              +\left(\frac{3}{2}\frac{\partial_z a}{a}+\frac{1}{2}\frac{\partial_z f_R}{f_R}
        \right),\\
  \mathcal{K}^{\dagger}&=&
     -\partial_z +\left(\frac{3}{2}\frac{\partial_z a}{a}
         +\frac{1}{2}\frac{\partial_z f_R}{f_R}  \right),
\end{eqnarray}
which indicates that there
is no graviton mode with $m^2<0$. For Eq. (\ref{Schrodinger}), the solution of the zero mode with $m=0$ is
\begin{eqnarray}
\label{zeromode}
~~~~~~~~~~~~~~~~\psi^{(0)}(z)=N_0 a^{3/2}(z)f_{R}^{1/2}(z).
\end{eqnarray}
It is easy to show that $\psi^{(0)}(z)$ is normalizable, i.e.,
\begin{eqnarray}
~~~~~~~~~~~~~~~~\int_{-\infty}^{\infty}|\psi^{(0)}(z)|^{2}dz<\infty,
\end{eqnarray}
which implies that the zero mode is localized on the brane.
Here we note that if $f_{R}(z)=1+2\alpha R(z)=0$ has solution $z=\pm z_0$, then the effective potentials $W$ would has singularities at $z=\pm z_0$ and the corresponding zero mode vanishes at that point.
{ We note here that by doing the KK reduction for the zero mode we will get the following relation between the effective four-dimensional Planck mass $M_{Pl}$ and the fundamental five-dimensional Planck mass $M_*$:
\begin{eqnarray}
~~~~~~~~~~~~~~~~~~~~~~M_{Pl}^2 \sim M_*^3 / k,
\end{eqnarray}
which results in the effective four-dimensional general relativity. Hence, it is natural to set $M_{Pl}$, $M_*$ and $k$ as the same scale (as did in RS-2 model)
\begin{eqnarray}
~~~~~~~~~~~~~~~~~~~~~~M_{Pl} \sim M_* \sim k,
\end{eqnarray}
so that there is no hierarchy between them.}

There is no analytical expression for the potential $W$ with respect to the coordinate $z$. However, we can express it in the coordinate $y$ as
\begin{eqnarray}
~~~\!\!\!\!W
&=&\frac{3}{4}\Big(\frac{a(y)a'(y)}{a(y)}\Big)^2
   +\frac{3}{2}\frac{a(y)a'^{2}(y)+a(y)^2a''(y)}{a(y)}\nonumber\\
&+&\frac{3\alpha a^{2}(y)a'(y)R'(y)}{a(y)(1+ 2\alpha R(y))}
   -\frac{\alpha^{2}a^{2}(y)R'^{2}(y)}{(1+ 2\alpha R(y))^2}  \nonumber\\
&+&
   \frac{\alpha\left[a(y)a'(y)R'(y)+ a^{2}(y)R''(y)\right]}
        {1+2\alpha R(y)}\nonumber\\
  &=&  \frac{k^2}{4}\textrm{sech}^{2B}(ky)
     \bigg( \! 15 B^2 \!-\!(2\!+\!3B)(4\!+\!5B)\textrm{sech}^{2}(k y)\nonumber\\
  & -&\frac{128B(2+5B)(1+16Bk^2\alpha)k^2\alpha}
                  {(1\!+\!8B(4\!+\!5B)k^2\alpha\!+\!(1\!-\!40B^2k^2\alpha)\cosh(2ky))^2}\nonumber\\
  &+&\!\!\!\frac{16(1+2B)(1+16Bk^2\alpha)}
                  {1\!+\!8B(4\!+\!5B)k^2\alpha\!+\!(1\!-\!40B^2k^2\alpha) \cosh(2ky)}
     \bigg).~~~ \label{Wzy}
\end{eqnarray}
From now on, we define dimensionless variables $\bar{\alpha}\equiv\alpha k^2$, $\bar{y}\equiv ky$, $\bar{z}\equiv kz$, $\bar{W}\equiv {W}/{k^2}$ and $\bar{m}\equiv {m}/{k}$. Then $\bar{z}$ is a function of just $\bar{y}$ and Eq.~(\ref{SchrodingerEq}) becomes
\begin{eqnarray}
~~~~~~~~~~~~\left[-\partial_{\bar{z}}^2+\bar{W}(\bar{z})\right]\psi(\bar{z})=\bar{m}^2\psi(\bar{z}),
\end{eqnarray}
where the effective potential is
\begin{eqnarray}
\bar{W}(\bar{z}(\bar{y}))\!\!&=&\!\!\frac{1}{4}\textrm{sech}^{2B}\bar{y}
   \Big[15 B^2 \!-\!(2\!+\!3B)(4\!+\!5B)\textrm{sech}^{2}\bar{y}\nonumber\\
  &-&\!\!\frac{128B(2+5B)(1+16B\bar{\alpha})\bar{\alpha}}
           {\big[1\!+\!8B(4\!+\!5B)\bar{\alpha}\!+\!(1\!-\!40B^2\bar{\alpha})\cosh(2\bar{y})\big]^2}\nonumber\\
  &+&\!\!\frac{16(1+2B)(1+16B\bar{\alpha})}
           {1\!+\!8B(4\!+\!5B)\bar{\alpha}\!+\!(1\!-\!40B^2\bar{\alpha})\cosh(2\bar{y})}
   \Big].~~~~
\end{eqnarray}
Now the parameter $k$ does not appear, which implies that we only need to consider different values of the dimensionless varieties $B$ and $\bar{\alpha}$ to search for resonant modes. For convenience, we remove the bars on all variables, which is equivalent to let $k=1$. So we have
\begin{eqnarray}
~~~~~~~~~~~~~~~~\alpha_1&=&-\frac{3}{32(1+4B)},\\
\alpha_s&=&-\frac{3+9B}{8(16+60B+49B^{2})},\\
\alpha_2&=&\frac{3}{8(8\!+\!16B\!+\!5B^2)}.
\end{eqnarray}

Note that $f_R = 1+2\alpha R(y)$ appears in the denominator of the expression of $W(z)$ (\ref{Wz}) or  $W(z(y))$ (\ref{Wzy}). For general relativity, we have $f_R = 1$ and so the effective potential $W(z(y))$ is always regular for a smooth solution of the thick brane. However, for the $f(R)$ gravity with $f(R) = R+\alpha R^2$, $f_R$ may be vanishing and the effective potential will be divergent, i.e.,
\begin{eqnarray}
~~~~1\!+\!2\alpha R(y)\!=\!1\!-\!8\alpha B[\!-\!2\!+\!(2\!+\!5B)\tanh^{2}(y)]\!=\!0 \label{criticalEq}
\end{eqnarray}
will lead to the singularities of the effective potential $W$. This will result in two $\delta$-like potential wells which are related to the appearance of the resonant KK modes of the tensor perturbations.

\begin{figure}
\begin{center}
\includegraphics[width=0.35\textwidth]{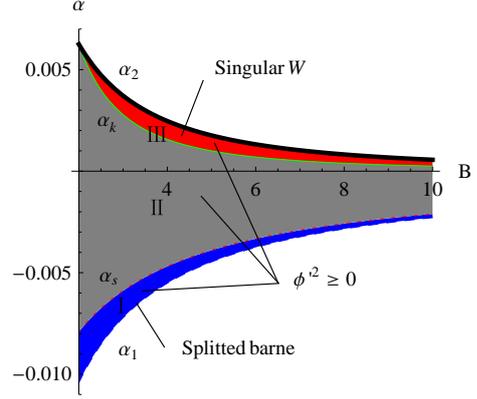}
\end{center}
\caption{The structure of the parameter space of $(\alpha,B)$ for the $f(R)$-brane model. Areas \MakeUppercase{\romannumeral1} ($\alpha_1<\alpha<\alpha_s$ for a fixed $B$), \MakeUppercase{\romannumeral2} ($\alpha_s<\alpha<\alpha_k$ for a fixed $B$), and \MakeUppercase{\romannumeral3} ($\alpha_k<\alpha<\alpha_2$ for a fixed $B$) correspond to $\phi'^2\geq0$. Area \MakeUppercase{\romannumeral1} corresponds to the splitted branes.  Area \MakeUppercase{\romannumeral3} corresponds to the singular $W$.}\label{Alpha}
\end{figure}

Now we analyze the parameter space of $(B, \alpha)$ where Eq.~(\ref{criticalEq}) has solutions. Since $1+2\alpha R(y)$ is an even function of $y$, we only consider the region $y\in(0,+\infty)$.
We first discuss the case of $\alpha>0$. It is clear that $1+2\alpha R(y)$ decreases with $y$ and $1+2\alpha R(0)=1+16\alpha B>0$. So the sufficient and necessary condition that Eq. (\ref{criticalEq}) has a solution is $1+2\alpha R(y\rightarrow+\infty)=1-40\alpha B^2<0$. Therefore, the relation between $\alpha$ and $B$ is
\begin{eqnarray}
~~~~~~~~~~~~~~~~~~~~~~~~\alpha>\frac{1}{40B^{2}}>0.
\end{eqnarray}
On the other hand, according to the constrain on $\alpha$ (\ref{ConstrainOnalpha}), we should have $\frac{1}{40B^{2}}<\alpha_{2}$, which results in $B>2$. Next, we discuss the case of $\alpha<0$, for which $1+2\alpha R(y)$ increases with $y$ and $1+2\alpha R({y\rightarrow+\infty})=1-40\alpha B^2>0$. So the sufficient and necessary condition that Eq. (\ref{criticalEq}) has a solution is $1+2\alpha R(0)=1+16\alpha B\leq0$. Therefore, the relation between $\alpha$ and $B$ is
\begin{eqnarray}
~~~~~~~~~~~~~~~~~~~~~~~~\alpha\leq-\frac{1}{16B}<0.
\end{eqnarray}
Furthermore, the constrain on $\alpha$ (\ref{ConstrainOnalpha}) requires $-\frac{1}{16B}\geq\alpha_{1}$, namely $B\leq-\frac{2}{5}$, which contradicts with $B>0$. So Eq. (\ref{criticalEq}) has no solution for negative $\alpha$. In conclusion, only when the constrain conditions
\begin{eqnarray}
~~~~~~~~~~~~B>2  ~~\textrm{and}~~\alpha_{k}\equiv\frac{1}{40B^{2}}<\alpha\leq\alpha_{2} \label{constrainConditionsForBandalpha}
\end{eqnarray}
are satisfied, does Eq.~(\ref{criticalEq}) have a solution. The relations among $\alpha_1, \alpha_s, \alpha_k, \alpha_2$ are shown in Fig.~\ref{Alpha}. Because of the monotony of it, $f_{R}$ (the derivative of $f(R)$ with respect of $R$) is negative in some range of $y$ as long as Eq.~(\ref{criticalEq}) has a solution. This will lead to the existence of ghosts.

\begin{figure*}
\begin{center}
\subfigure[$W(y,\alpha)$ with $B=2$]{\label{W2}
\includegraphics[width=0.23\textwidth]{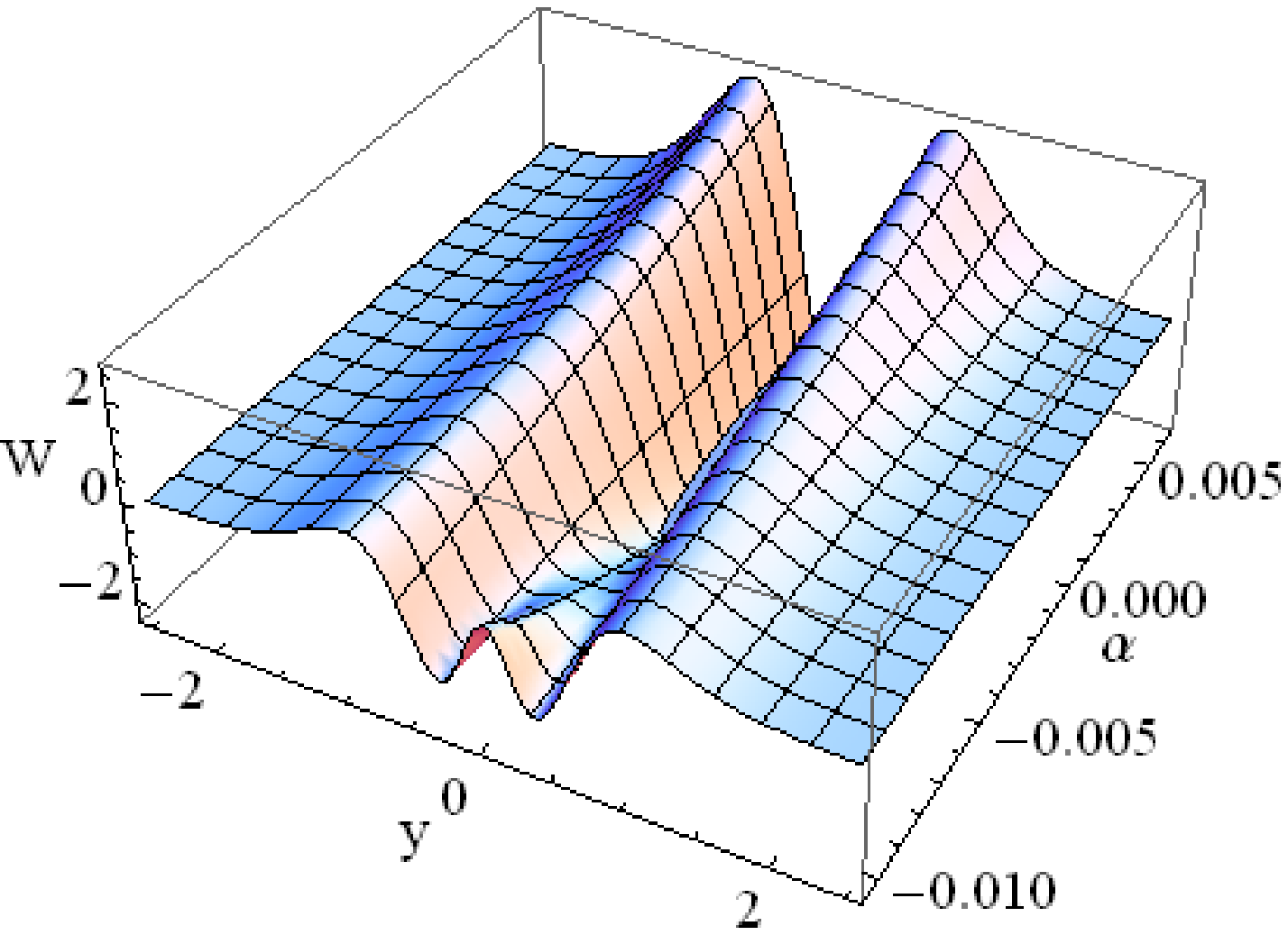}}
\subfigure[$W_s(y,\alpha)$ with $B=2$]{\label{WS2}
\includegraphics[width=0.23\textwidth]{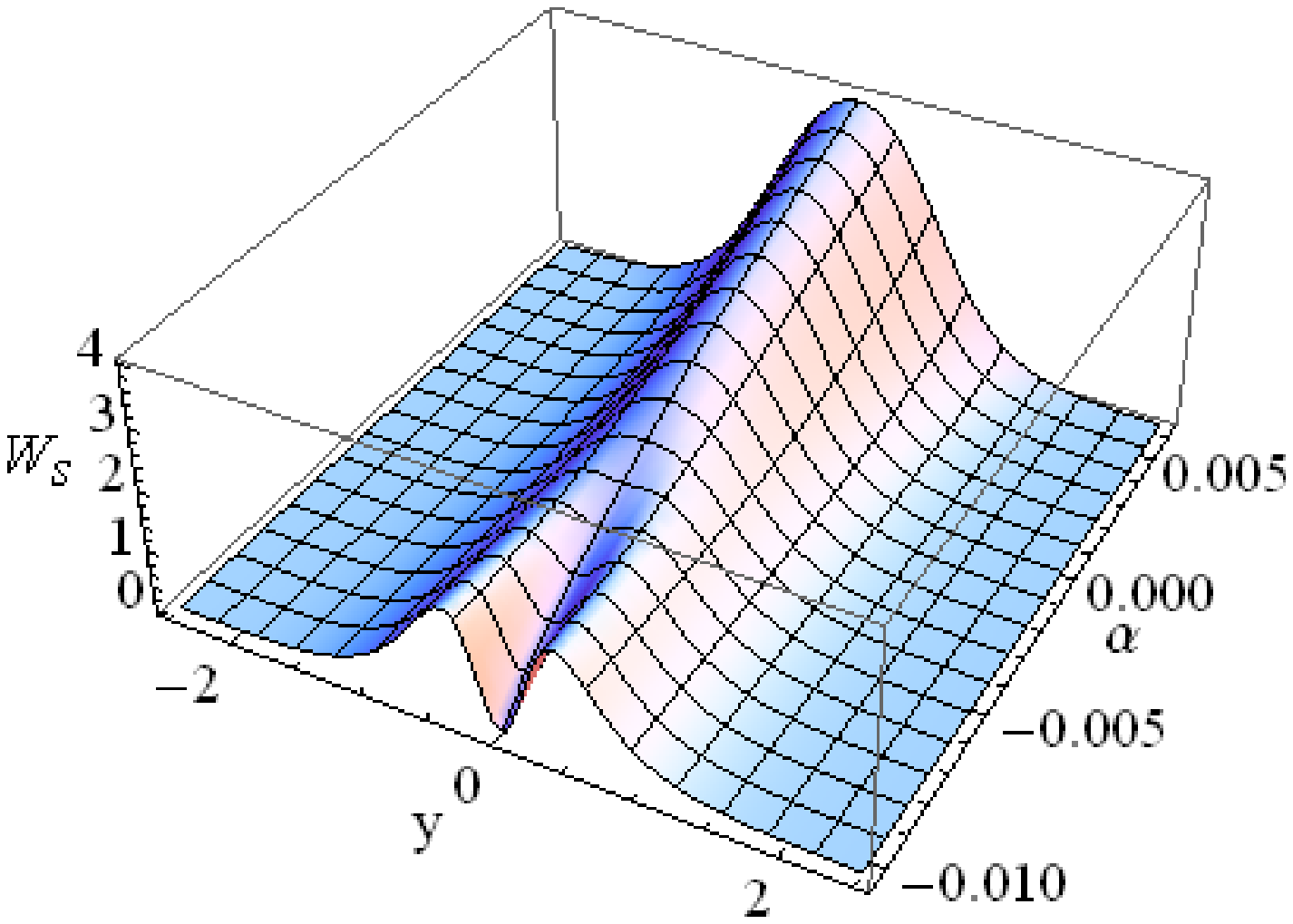}} 
\subfigure[$W(y,\alpha)$ with $B=4$]{\label{W4}
\includegraphics[width=0.23\textwidth]{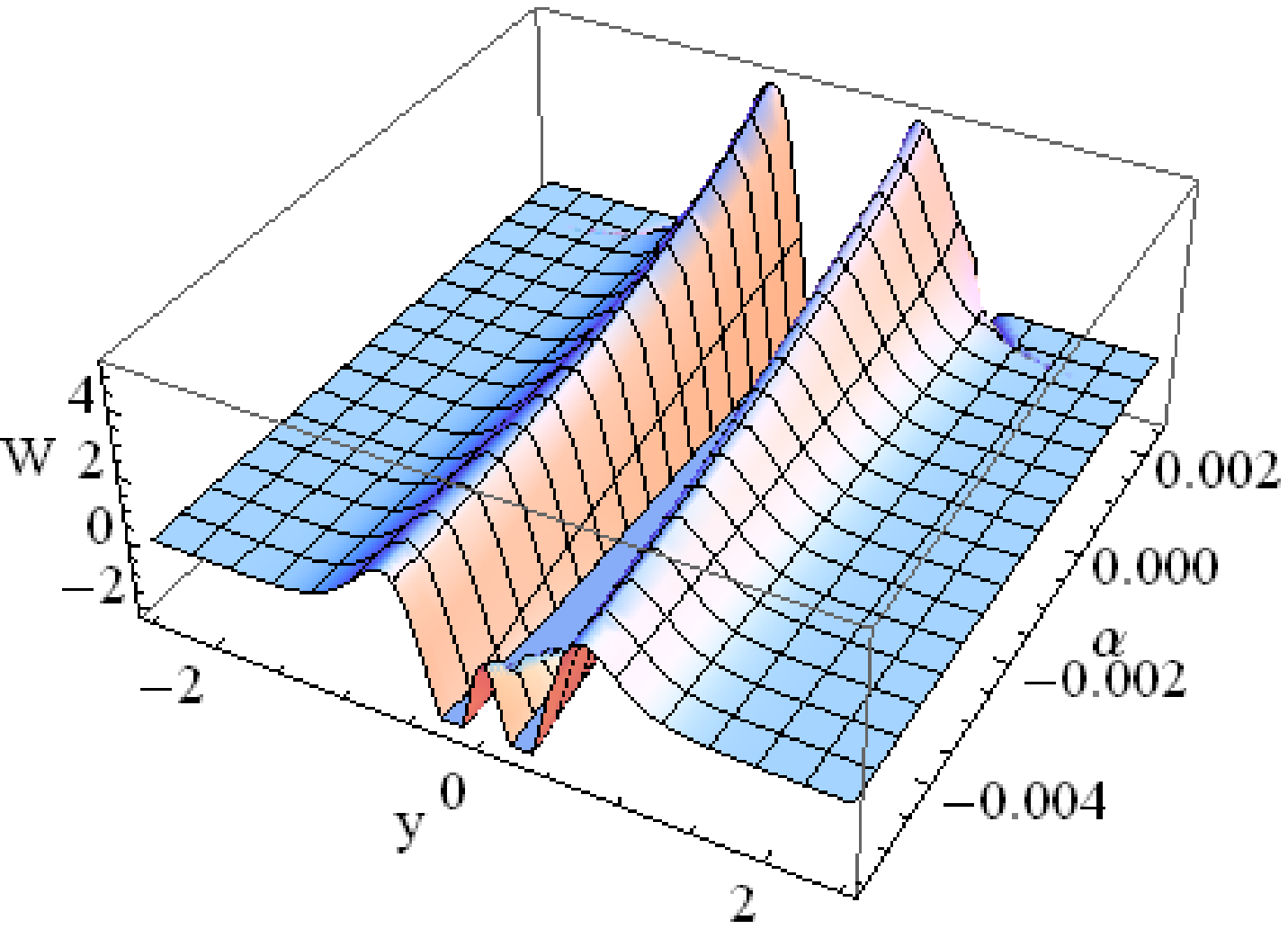}}
\subfigure[$W_s(y,\alpha)$ with $B=4$]{\label{WS4}
\includegraphics[width=0.23\textwidth]{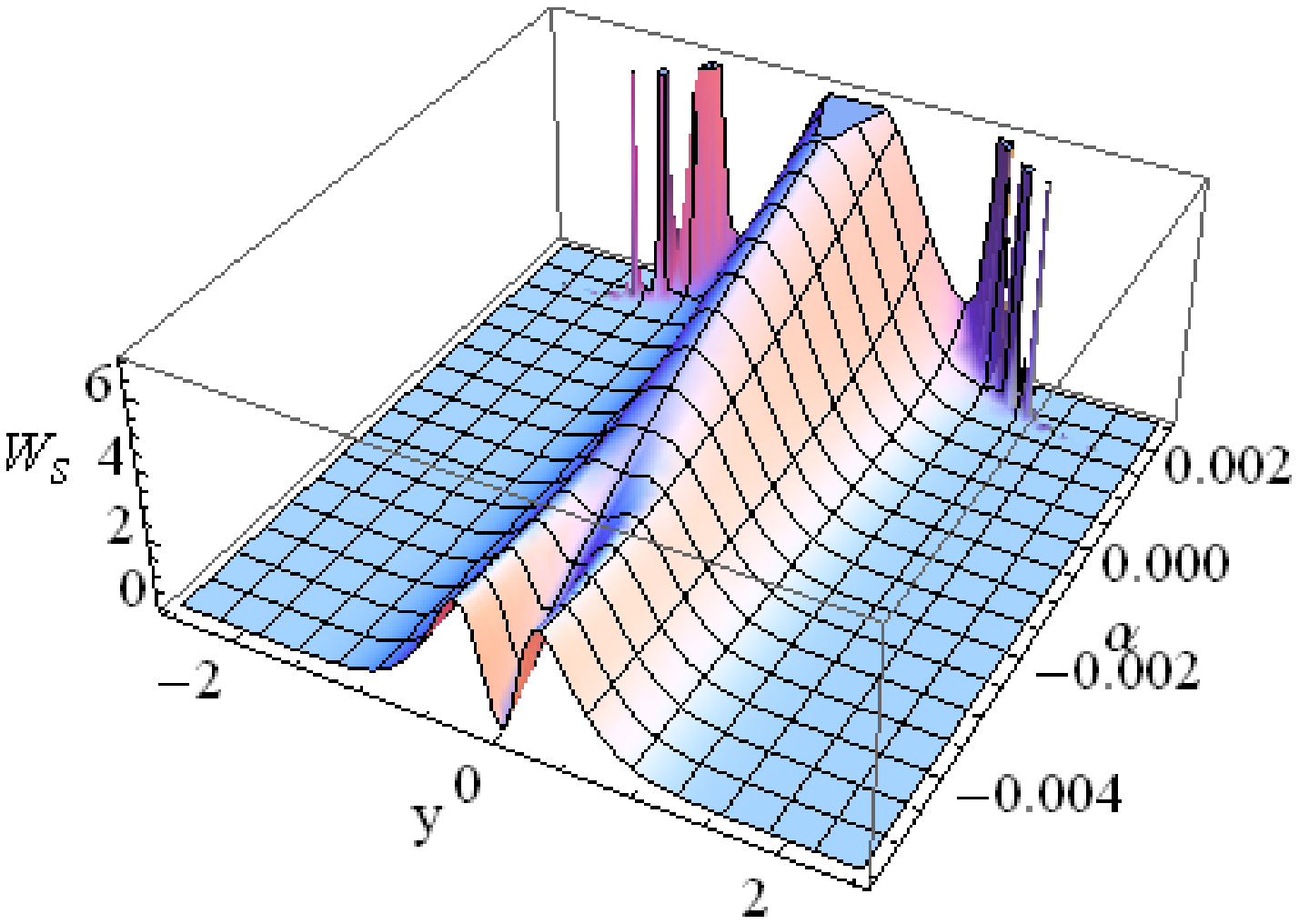}}
\end{center}
\caption{The effective potential $W$ and the dual one $W_s$ as functions of $y$ and $\alpha$ ($\alpha_{1}\leq\alpha\leq\alpha_{2}$). The parameters are set to $B=2$ (a, b), and $B=4$ (c, d).}\label{WandWs}
\end{figure*}

In order to judge whether there are resonant modes, we consider the  partner equation of the Schr\"odinger-like equation (\ref{SchrodingerEq2}): $\mathcal{K}^{\dagger}\mathcal{K}\psi(z) =m^2\psi(z)$, for which
the corresponding potential is given by
\begin{eqnarray}
~~~~~~~~W_{s}(z)&=&\frac{15}{4}\frac{(\partial_{z}a)^2}{a^2}
   -\frac{3}{2}\frac{\partial_{z}^{2}a}{a}+\frac{3}{2}\frac{\partial_{z}a \partial_{z}f_R}{a f_R}
  \nonumber \\
  &&\!\!+\frac{3}{4}\frac{(\partial_{z}f_R)^2}{f_R^2}-\frac{1}{2}\frac{\partial_{z}^{2}f_R}{f_R}.
\end{eqnarray}
Similar to $W(z)$, there is no analytical expression for $W_{s}(z)$. In the $y$ coordinate, we have
\begin{eqnarray}
\label{WS}
\!\!\!\!&&W_{s}(z(y))=\frac{1}{4}\textrm{sech}^{2B}y
     \Big[3B^2-(2B+3B^2)\textrm{sech}^{2}y\nonumber\\
   &&~~~~+\frac{384B(2+5B)(1+16B\alpha)\alpha}
                 {\big[1+8B(4+5B)\alpha
                  +(1-40B^2\alpha)\cosh(2y)\big]^2}\nonumber\\
   &&~~~~+\frac{16B-128B(2+3B)\alpha}{1+8B(4+5B)\alpha+(1-40B^2\alpha)\cosh(2y)}
    \Big].
\end{eqnarray}
The structure of $W$ and $W_{s}$ is shown in Fig. \ref{WandWs}.

Because $\frac{dz}{dy}=a(y)$ is positive, the shape of $W_{s}(z)$ is the same with that of $W_{s}(z(y))$. Thus, in order to estimate whether there are resonant modes we just need to consider the shape of $W_{s}(z(y))$. The appearance of ``quasiwell" of the dual potential may lead to the resonances of the graviton KK modes. Noticing
\begin{eqnarray}
~~~~~~~~~~~~~~~W_{s}(0)=-W(0)>0,\\
0<W_{s}|_{z\rightarrow\pm\infty}\rightarrow0,\\
0<W|_{z\rightarrow\pm\infty}\rightarrow0,
\end{eqnarray}
it is convenient to judge whether there are resonant modes for Eq.~(\ref{SchrodingerEq}) from $W_{s}$. Solving $\partial_{y}^{2}W|_{y=0}=0$ and $\partial_{y}^{2}W_{s}|_{y=0}=0$ respectively results in
\begin{eqnarray}
\alpha&=&\frac{\!-\!8 \!-\! 36 B \!-\! 46 B^2 \!+\! (2 \!+\! 5 B)\sqrt{(16\!+\!
    52B \!+\! 67 B^2)}}{16 B (2 \!+\! 3 B) (4 \!+\! 7 B)}\equiv\alpha', \nonumber\\
\alpha&=&\frac{8 \!+\! 24B \!-\! 2B^2 \!-\! (2 \!+\! 5 B)\sqrt{(16 \!+\! 16B \!+\! 13 B^2)}}{
 16 B^2 (38 \!+\! 107 B)}\equiv\alpha''. \nonumber
\end{eqnarray}
This implies that both the effective potential and the corresponding dual one have an interval structure in the range of $\alpha_{1}<\alpha<\alpha'$ and the range of $\alpha_{1}<\alpha<\alpha''$, respectively (see Fig.~\ref{WandWs}). Furthermore, it can be seen from Fig.~\ref{WandWs} that there is also another interesting quasiwell with singularity for large $B$ and $\alpha$ satisfying the constrain conditions (\ref{constrainConditionsForBandalpha}).

To get the numerical solution of Eq. (\ref{Schrodinger}), we impose the following conditions:
\begin{eqnarray}
~~~~~~~~~~~~~\psi_{\rm{even}}(0)\!\!&=&\!\!1, ~~~\partial_{z}\psi_{\rm{even}}(0)=0;\\
\psi_{\rm{odd}}(0)\!\!&=&\!\!0, ~~~~\partial_{z}\psi_{\rm{odd}}(0)=1.
\end{eqnarray}
Here $\psi_{\rm{even}}$ and $\psi_{\rm{odd}}$ denote the even
and odd parity modes of $\psi(z)$, respectively. The solution under this imposition does not affect the relative probability defined below.

The function $|\psi(z)|^{2}$ can be interpreted as the probability of finding the massive KK modes at the
position $z$ along extra dimension \cite{LiuYangZhaoFuDuan2009}.
In order to find the resonant modes, we definite the relative probability \cite{LiuYangZhaoFuDuan2009}
\begin{eqnarray}
~~~~~~~~~~~~~~~~~~P(m^{2})=\frac{\int^{z_{b}}_{-z_{b}}|\psi(z)|^{2}dz}{\int^{z_{max}}_{-z_{max}}|\psi(z)|^{2}dz},
\end{eqnarray}
where $2z_{b}$ is approximately the width of the brane, and $z_{max}$ is taken as $z_{max}=10z_{b}$. It is clear that large relative probabilities $P(m^{2})$ of finding massive KK modes within a narrow range $-z_b < z < z_b$ around the brane location indicate the existence of resonant modes.

From Fig. \ref{WandWs}, we can see that there is a quasiwell when $\alpha\rightarrow\alpha_1$, which is consistent with what we analyse. So, it seems that we may find resonant modes for small $\alpha$. When $\alpha\rightarrow\alpha_2$ and $B\ge4$ which satisfies the condition (\ref{constrainConditionsForBandalpha}), the effective potential has two $\delta$-like potential wells and there are  the resonant modes found.

\begin{figure*}
\begin{center}
\subfigure[$B=2$, $\alpha=\alpha_{1},\alpha_{s},0$]{\label{WB21}
\includegraphics[width=0.3\textwidth]{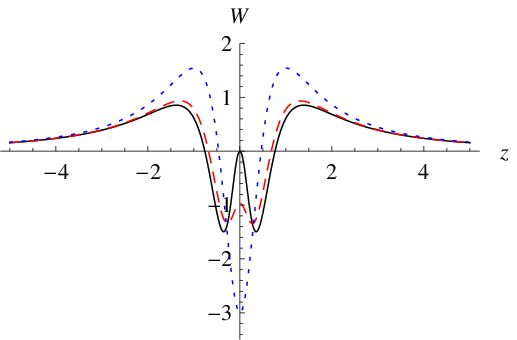}}
\subfigure[$B=2$, $\alpha=\alpha_{2}$]{\label{WB22}
\includegraphics[width=0.3\textwidth]{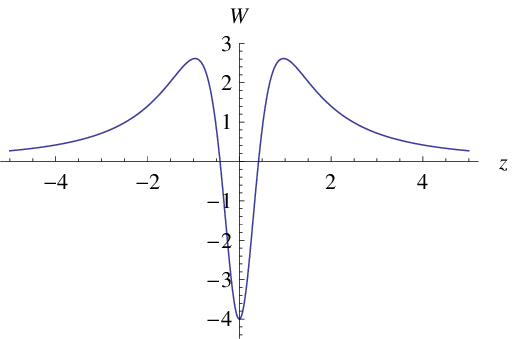}}
\subfigure[$B=4$, $\alpha=\alpha_{1},\alpha_{s},0$]{\label{WB41}
\includegraphics[width=0.3\textwidth]{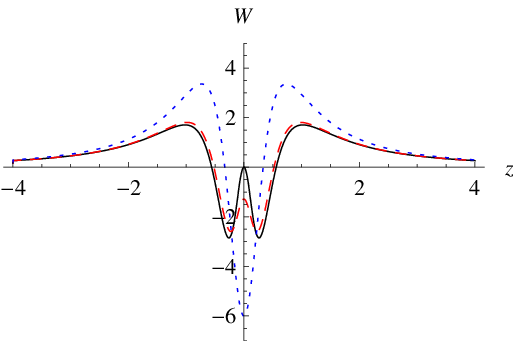}}
\subfigure[$B=4$, $\alpha=\alpha_{k}$]{\label{WB42}
\includegraphics[width=0.3\textwidth]{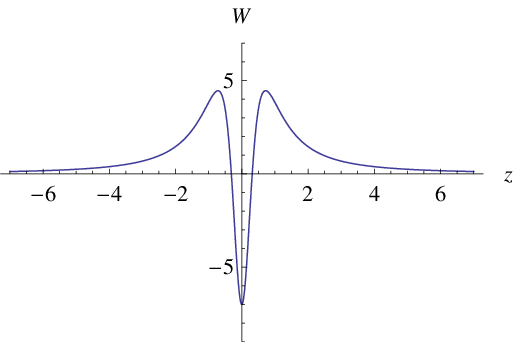}}
\subfigure[$B=4$, $\alpha=\frac{\alpha_{k}+\alpha_{2}}{2}$]{\label{WB43}
\includegraphics[width=0.3\textwidth]{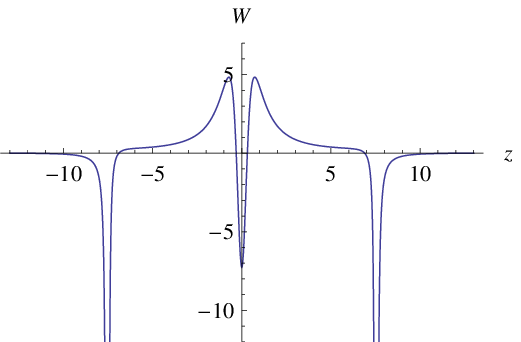}}
\subfigure[$B=4$, $\alpha=\alpha_{2}$]{\label{WB44}
\includegraphics[width=0.3\textwidth]{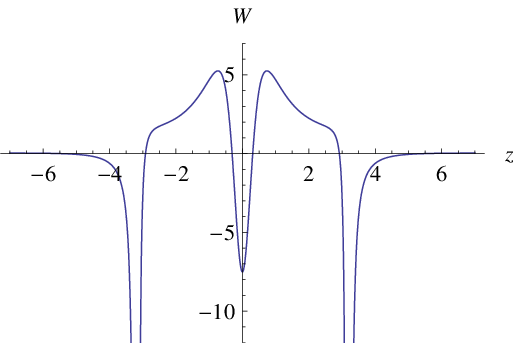}}
\end{center}
\caption{The shapes of the effective potential $W(z)$. The parameters are set to $B=2$ (a,b) and $B=4$ (c,d), $\alpha=\alpha_{1}$ (black line), $\alpha=\alpha_{s}$ (dashed red line), and $\alpha=0$ (dotted blue line) (a,c)}.
\label{WB}
\end{figure*}

\begin{figure*}
\begin{center}
\subfigure[$B=2$, $\alpha=\alpha_{1}$]{\label{ResonanceB2a1}
\includegraphics[width=0.23\textwidth]{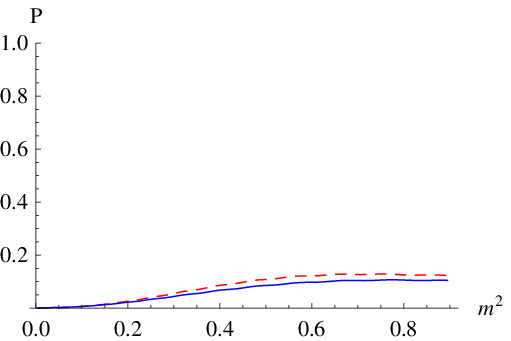}}
\subfigure[$B=2$, $\alpha=\alpha_{s}$]{\label{ResonanceB2a2}
\includegraphics[width=0.23\textwidth]{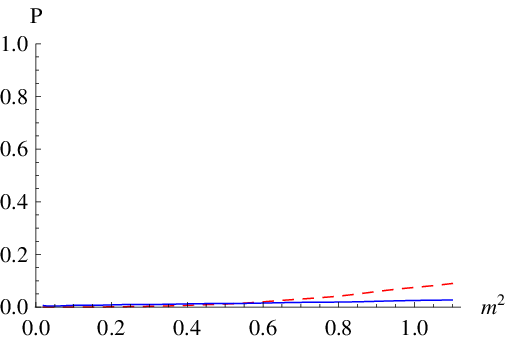}} 
\subfigure[$B=2$, $\alpha=0$]{\label{ResonanceB2a3}
\includegraphics[width=0.23\textwidth]{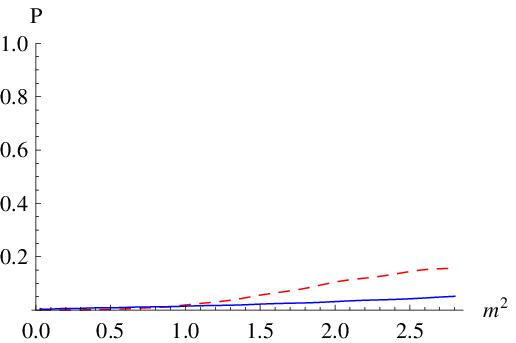}}
\subfigure[$B=2$, $\alpha=\alpha_{2}$]{\label{ResonanceB2a4}
\includegraphics[width=0.23\textwidth]{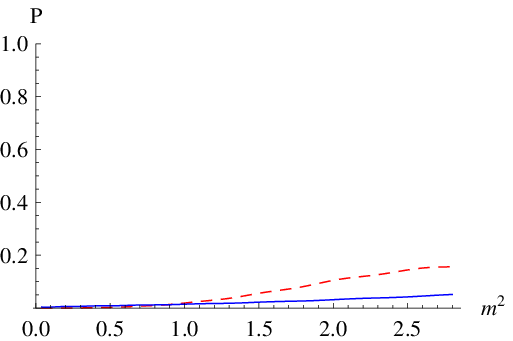}}
\end{center}
\caption{The relative probability $P(m^2)$ of the KK modes for Eq.~(\ref{Schrodinger}) with $B=2$. The dashed red and solid blue lines are for odd and even parity KK modes, respectively.} \label{ResonanceB2}
\end{figure*}

Firstly, we discuss the case of $B\leq2$. When $\alpha$ approaches to $\alpha_{1}$, the dual potential $W_{s}(z)$ has a quasiwell in the middle, so resonant modes of Eq.~(\ref{Schrodinger}) may appear. We use the numerical method to search for resonant modes for $B\leq2$ but no resonant mode is found. The reason is that the quasiwell is not deep or wide enough. The potential $W(z)$ for $B=2$ is shown in Figs.~\ref{WB21} and \ref{WB22}. The relations between the relative probability $P$ and the mass square $m^{2}$ for $B=2$ are shown in Fig.~\ref{ResonanceB2}.

\begin{figure*}
\begin{center}
\subfigure[$B=4$, $\alpha=\alpha_{1}$]{\label{ResonanceB4a1}
\includegraphics[width=0.3\textwidth]{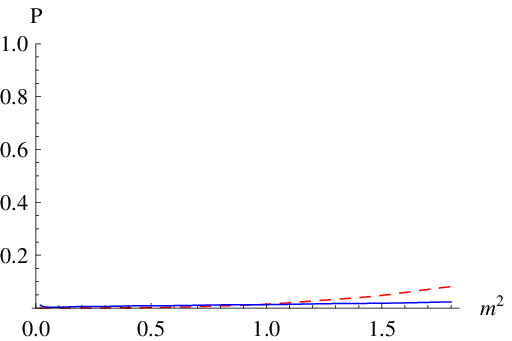}}
\subfigure[$B=4$, $\alpha=\alpha_{s}$]{\label{ResonanceB4a2}
\includegraphics[width=0.3\textwidth]{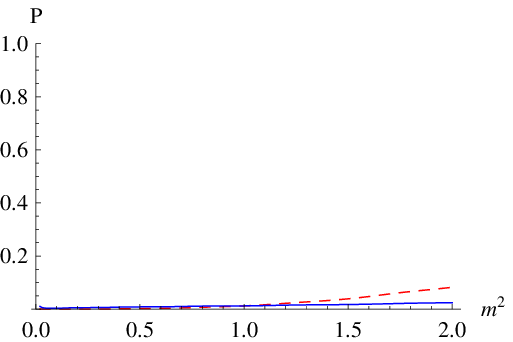}}
\subfigure[$B=4$, $\alpha=0$]{\label{ResonanceB4a3}
\includegraphics[width=0.3\textwidth]{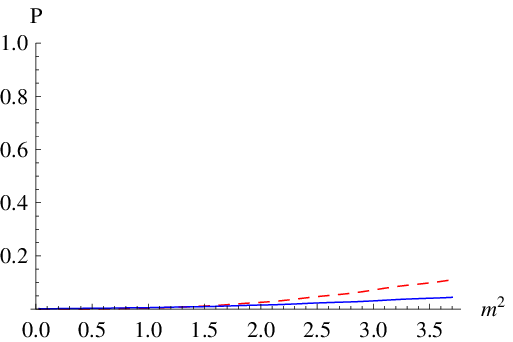}}
\subfigure[$B=4$, $\alpha=\alpha_{k}$]{\label{ResonanceB4a4}
\includegraphics[width=0.3\textwidth]{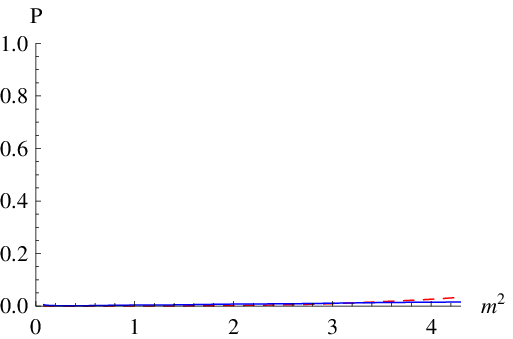}}
\subfigure[$B=4$, $\alpha=\frac{\alpha_{k}+\alpha_{2}}{2}$]{\label{ResonanceB4a5}
\includegraphics[width=0.3\textwidth]{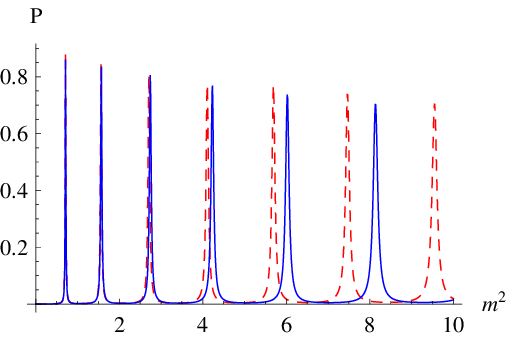}}
\subfigure[$B=4$, $\alpha=\alpha_{2}$]{\label{ResonanceB4a6}
\includegraphics[width=0.3\textwidth]{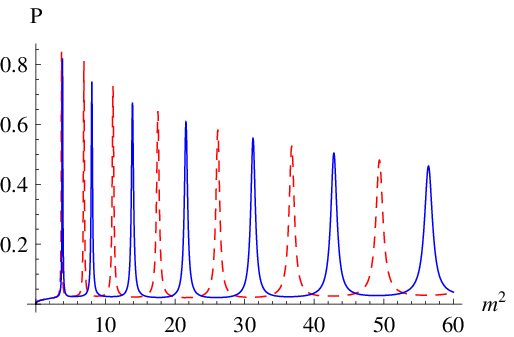}}
\end{center}
\caption{The relative probability $P(m^2)$ of the KK modes for Eq. (\ref{Schrodinger}) with $B=4$ and $k=1$. The dashed red and solid blue lines are for odd and even parity KK modes, respectively.} \label{ResonanceB4}
\end{figure*}

\begin{figure*}
\begin{center}
\includegraphics[width=0.3\textwidth]{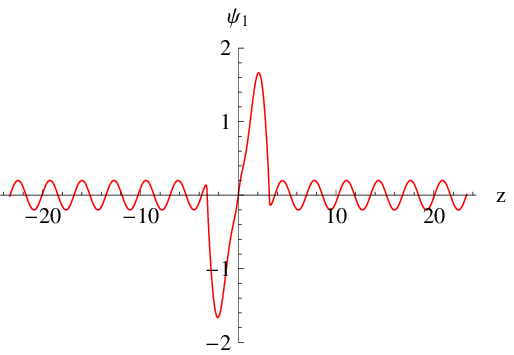}
\includegraphics[width=0.3\textwidth]{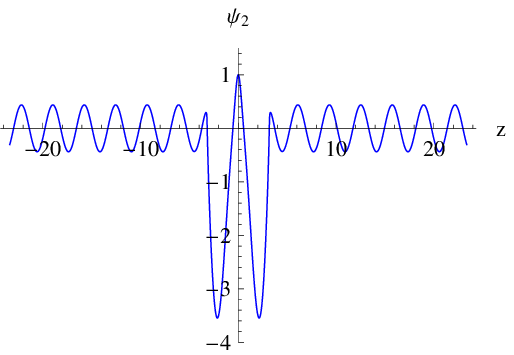}
\includegraphics[width=0.3\textwidth]{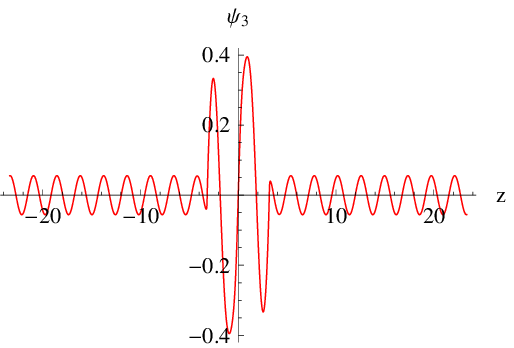}\\ 
\includegraphics[width=0.3\textwidth]{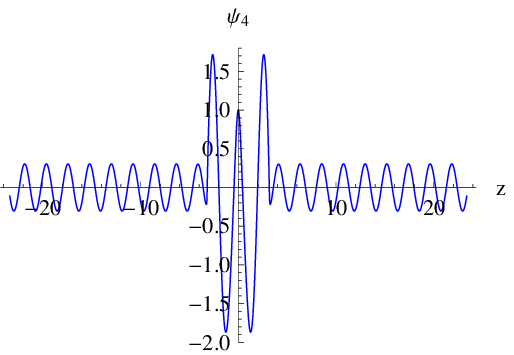}
\includegraphics[width=0.3\textwidth]{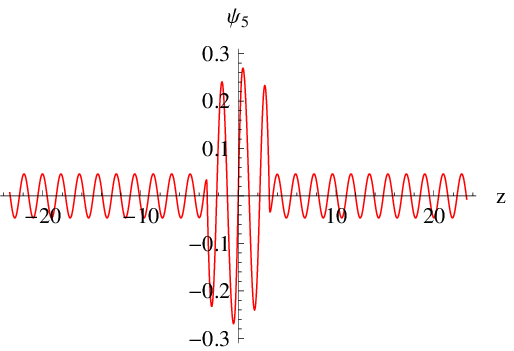}
\includegraphics[width=0.3\textwidth]{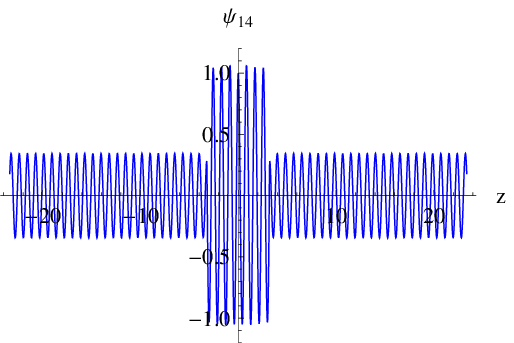}
\end{center}
\caption{The resonant wave function $\psi_n(z)$ for Eq. (\ref{Schrodinger}) with $B=4$, $k=1$ and $\alpha=\alpha_{2}$. The red and solid blue lines are for odd and even parity KK modes, respectively. The coordinate axis $\phi_{n}$ denotes wave function for the $n$-th resonant mode.} \label{wavefunction}
\end{figure*}

Secondly, we discuss the case of $B>2$. Similar to the case of $B\leq2$, there is a quasiwell in the middle of the dual potential $W_{s}(z)$, but no resonant modes are found when $\alpha_1\leq\alpha\leq\alpha_k$ (see Figs.~\ref{WB41}, \ref{WB42}, and Figs.~\ref{ResonanceB4a1}-\ref{ResonanceB4a4}). When $\alpha_k<\alpha\leq\alpha_{2}$, the potential $W(z)$, which has $Z_{2}$ symmetry and is shown in Figs.~\ref{WB43} and \ref{WB44}, will have two $\delta$-like potential wells. A series of resonant modes come forth (see Figs.~\ref{ResonanceB4a5} and \ref{ResonanceB4a6}) and their wave functions are shown in Fig.~\ref{wavefunction} for $B=4$.

\begin{figure*}
\begin{center}
\subfigure[$B=4$, $\alpha=0$]{\label{phino}
\includegraphics[width=0.3\textwidth]{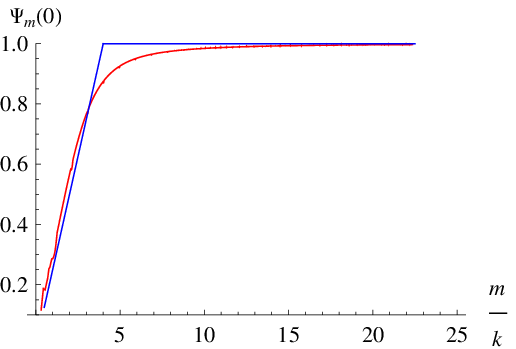}}
\subfigure[$B=4$, $\alpha=\alpha_{2}$]{\label{phiyes}
\includegraphics[width=0.3\textwidth]{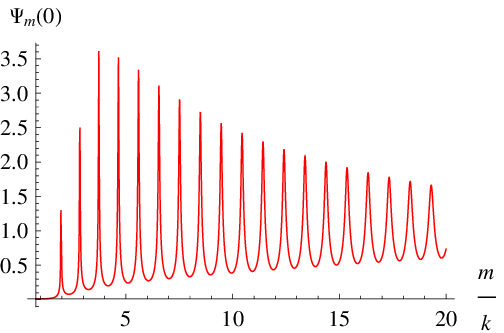}}
\end{center}
\caption{$\psi_m(0)$ for resonant wave function $\psi(z)$ with $B=4$, $\alpha=0$ and $B=4$, $\alpha=\alpha_{2}$. The red lines are corresponding numerical result and the blue line is an approximate linear fitting.} \label{phi}
\end{figure*}

The last point of interest for this section is to calculate the contribution of the massive graviton KK modes including the resonances to Newton's law. The correction to Newton's law between two mass points $m_1$ and $m_2$ localized at $z=0$, with a distance $r$ from the KK modes is given by \cite{Lykken1999nb,Csaki:2000fc}:
{\begin{eqnarray}
\label{UR}
~~~~~~U(r)\thicksim G_N\frac{m_1m_2}r\Big[1\!+\!\int_{0}^\infty \frac{d m}{k}
e^{-mr}\psi_m^2(0)\Big],
\end{eqnarray}
where $\psi_m(z)$ is normalized such that $\psi_m(z)|_{z\rightarrow\infty}\simeq\cos(z)$,  and the effective four-dimensional Newton's constant $G_N$ is given by
\begin{eqnarray}
~~~~~~~~~~~~~~~~~~~~~~~~~~G_N \sim M^{-2}_{Pl}.
\end{eqnarray}
Next, we first calculate the expression of $\psi_m(0)$.} As examples, we just discuss two representative cases: $B=4,\alpha=0$ and $B=4 ,\alpha=\alpha_2$. The corresponding numerical results are shown in Fig.~\ref{phi}. As we can see from Figs.~\ref{phino} and~\ref{phiyes}, the trend of {$\psi_m(0)$} with the $m$ changing is analogous to relative probability $P(m^2)$ (see Figs.~\ref{ResonanceB4a3} and \ref{ResonanceB4a6}). Secondly, in order to do the integration in Eq.~(\ref{UR}) we need to obtain analytical expression of {$\psi_m(0)$} by using some approximate methods. One of the simplest methods is that we can simulate the original function with a few simple linear functions. For example, for the case of $B=4,\alpha=0$, the original function can be divided into two parts (see the blue line in Fig.~\ref{phino}):
{\begin{eqnarray}
~~~~~~~~~~~~~~\psi_m(0)&=&\frac{m}{4 k},~~~~~~~~~~~~0\leq m\leq 4k\nonumber\\
\psi_m(0)&=&1,~~~~~~~~~~~~~~~~m>4k,
\end{eqnarray}
Note that $\psi_m(0)$ is dimensionless. Substituting this result into Eq.~(\ref{UR}), we get the approximative expression for the gravity potential $U(r)$:
\begin{eqnarray}
U(r)=G_N\frac{m_1m_2}r\left(1+\frac{e^{-4 k r} \left(-4 k r+e^{4 k r}-1\right)}{8 (kr)^3}\right).
\end{eqnarray}
We expand $U(r)$ in terms of $kr$ under two extreme situations:
\begin{eqnarray}
&U(r\ll 1/k)&\!\sim\!\frac{G_Nm_1m_2}{r}\left(1\!+\!\frac{1}{k r}\!-\!
\frac{8}{3}\!+\!4 kr\!
   +\!\mathcal{O}[(kr)^2]\right),\nonumber\\
&U(r\gg 1/k)&\!\sim\! \frac{G_Nm_1m_2}{r}\left(1+\!\frac{1}{8(kr)^3}
    \!+\!\mathcal{O}\left[\frac{1}{(kr)^4}\right]\right).
\end{eqnarray}
For the case without resonance, the correction to Newton's law has the following obvious characteristic. The main correction occurs at a short-distance $r$ of $r<1/k\sim\times10^{-33}$cm and it is proportional to $\frac{1}{r^2}$. So, in the case of the current accuracy of the experiment~\cite{Long:2002wn}, such correction is unobservable and undetectable in this brane model. At a large-distance ($r\gg 1/k$), the correction can be neglected.} For the case of $B=4,~\alpha=\alpha_2$ with multiple resonances, the analysis and method are similar. Because the value of {$\psi_m(0)$} inevitably trends to be $1$ when the parameter $m$ approaches to infinite, the leading correction to Newton's law at a large-distance $r\gg 1/k$ is also proportional to $\frac{1}{r^4}$. At a short-distance $r<1/k$, the leading term is $\frac{G_Nm_1m_2}{kr^2}$. The effect of the resonant modes occurs  at a few Planck lengths. It is very difficult to give the expression of this correction.

\section{Conclusion}

\label{Conclusion}

In this paper, we investigated localization and resonant KK modes of the tensor fluctuation for the $f(R)$-brane model with $f(R)=R+\alpha R^{2}$. This investigation was based on the interesting analytical brane solution found in Ref.~\cite{BazeiaLobaoMenezesPetrovSilva2014}, where the warp factor was given by $\textrm{e}^{A(y)}=\textrm{sech}^{B}(ky)$ and the parameter $\alpha$ is constrained as $\alpha_{1}\leq\alpha\leq \alpha_{2}$ (see Eq. (\ref{ConstrainOnalpha})).

It was found that, when $\alpha_{1}\leq\alpha\leq\alpha_{s}(<0)$, the brane has an internal structure, but the effective potential $W(z)$ for the KK modes of the tensor fluctuation has only a simple structure (volcano-like potential with a single or double well) and no resonances were found. The reason is that the quasiwell of the effective potential is not deep enough.
This is different from the brane models in general relativity, for which there are usually resonant modes when the brane has an internal structure \cite{GuoLiuZhaoChen2012,XieYangZhao2013,CruzSousaMalufAlmeida2014}.

However, the effective potential may have a rich structure with singularities and it will support a series of resonant KK modes when $B$ is large enough ($B>2$) and $\alpha>0$, although the brane has no inner structure anymore. The reason is that the effective potential for the $f(R)$-brane model is decided by both the warp factor and the function $f(R)$. It was found that, when $B>2$ and $\frac{1}{40B^{2}k^{2}}<\alpha\leq\alpha_{2}$, the effective potential has two $\delta$-like potential wells and there are a series of resonant modes on the brane.

{ The existence of resonant modes contributes a correction to the Newtonian graviton potential at short distance of the Planck scale.} But $f_R$ is negative in some range so long as the effective potential has singularities because of the monotony of $f_R$. However, this anomaly will result in the appearance of ghosts. So we hope to construct a $f(R)$-brane model with positive $f_R$ and graviton resonances in the future.

\begin{acknowledgements}
This work was supported by the National Natural Science Foundation of China (Grant No. 11375075), and the Fundamental Research Funds for the Central Universities (Grant No. lzujbky-2015-jl01).
\end{acknowledgements}



\begin{thebibliography}{38}%
\providecommand{\url}[1]{{#1}}
\providecommand{\urlprefix}{URL }
\expandafter\ifx\csname urlstyle\endcsname\relax
  \providecommand{\doi}[1]{DOI \discretionary{}{}{}#1}\else
  \providecommand{\doi}{DOI \discretionary{}{}{}\begingroup
  \urlstyle{rm}\Url}\fi

\bibitem{Antoniadis1990}
I. Antoniadis, Phys. Lett. B, \textbf{246}, 377 (1990).

\bibitem{AntoniadisArkani-HamedDimopoulosDvali1998}
I. Antoniadis, N. Arkani-Hamed, S. Dimopoulos, and G. Dvali, Phys. Lett. B, \textbf{436}, 257 (1998), arXiv:hep-ph/9804398[hep-ph].

\bibitem{Arkani-HamedDimopoulosDvali1998a}
N. Arkani-Hamed, S. Dimopoulos, and G. Dvali, Phys. Lett. B, 429, 263 (1998), arXiv:hep-ph/9803315[hep-ph].

\bibitem{RandallSundrum1999}
L. Randall and R. Sundrum, Phys. Rev. Lett., \textbf{83}, 3370 (1999), arXiv:hep-ph/9905221.

\bibitem{RandallSundrum1999a}
L. Randall and R. Sundrum, Phys. Rev. Lett., \textbf{83}, 4690
(1999), arXiv:hep-th/9906064.

\bibitem{Gremm2000}
M. Gremm, Phys. Lett. B, \textbf{478}, 434 (2000), arXiv: hep-th/9912060.

\bibitem{KehagiasTamvakis2001}
A. Kehagias and K. Tamvakis, Phys. Lett. B, \textbf{504}, 38 (2001), arXiv:hep-th/0010112[hep-th].

\bibitem{CsakiErlichHollowoodShirman2000}
C. Csaki, J. Erlich, T. J. Hollowood, and Y. Shirman, Nucl. Phys. B, \textbf{581}, 309 (2000), arXiv:hep-th/0001033.

\bibitem{Hinterbichler2012}
K. Hinterbichler, Rev. Mod. Phys., \textbf{84}, 671 (2012), arXiv:1105.3735[hep-th].

\bibitem{Rham2014}
C. de Rham, (2014), arXiv:1401.4173[hep-th].

\bibitem{GuoLiuZhaoChen2012}
H. Guo, Y.-X. Liu, Z.-H. Zhao, and F.-W. Chen, Phys. Rev. D, \textbf{85}, 124033 (2012), arXiv:1106.5216[hep-th].

\bibitem{XieYangZhao2013}
Q.-Y. Xie, J. Yang, and L. Zhao, Phys. Rev. D, \textbf{88}, 105014 (2013), arXiv:1310.4585[hep-th].

\bibitem{CruzSousaMalufAlmeida2014}
W. Cruz, L. Sousa, R. Maluf, and C. Almeida, Phys. Lett. B, \textbf{730}, 314 (2014), arXiv:1310.4085[hep-th].

\bibitem{ZhongLiuZhao2014}
Y. Zhong, Y.-X. Liu, and Z.-H. Zhao, (2014), arXiv: 1404.2666[hep-th].

\bibitem{SotiriouFaraoni2010}
T. P. Sotiriou and V. Faraoni, Rev. Mod. Phys., \textbf{82}, 451 (2010), arXiv:0805.1726[gr-qc].

\bibitem{DeTsujikawa2010}
A. De Felice and S. Tsujikawa, Living Rev. Rel., \textbf{13}, 3 (2010), arXiv:1002.4928[gr-qc].

\bibitem{NojiriOdintsov2011}
S. Nojiri and S. D. Odintsov, Phys. Rept., \textbf{505}, 59 (2011), arXiv:1011.0544[gr-qc].

\bibitem{ZhongLiuYang2011}
Y. Zhong, Y.-X. Liu, and K. Yang, Phys. Lett. B, \textbf{699}, 398 (2011), arXiv:1010.3478[hep-th].

\bibitem{ParryPichlerDeeg2005}
M. Parry, S. Pichler, and D. Deeg, JCAP, \textbf{0504}, 014 (2005), arXiv:hep-ph/0502048.

\bibitem{AfonsoBazeiaMenezesPetrov2007}
V. I. Afonso, D. Bazeia, R. Menezes, and A. Y. Petrov, Phys. Lett. B, \textbf{658}, 71 (2007), arXiv:0710.3790[hep-th].

\bibitem{DeruelleSasakiSendouda2008}
N. Deruelle, M. Sasaki, and Y. Sendouda, Prog. Theor. Phys., \textbf{119}, 237 (2008), arXiv:0711.1150[gr-qc].

\bibitem{BalcerzakDabrowski2008}
A. Balcerzak and M. P. Dabrowski, Phys. Rev. D, \textbf{77}, 023524 (2008), arXiv:0710.3670[hep-th].

\bibitem{DzhunushalievFolomeevKleihausKunz2010}
V. Dzhunushaliev, V. Folomeev, B. Kleihaus, and J. Kunz, J. High Energy Phys., \textbf{04}, 130 (2010), arXiv:
0912.2812[gr-qc].

\bibitem{LiuZhongZhaoLi2011}
Y.-X. Liu, Y. Zhong, Z.-H. Zhao, and H.-T. Li, J. High Energy Phys., \textbf{06}, 135 (2011), arXiv:1104.3188[hep-th].

\bibitem{HoffDias2011}
J. Hoff da Silva and M. Dias, Phys. Rev. D, \textbf{84}, 066011 (2011), arXiv:1107.2017[hep-th].

\bibitem{LiuLuWang2012}
H. Liu, H. Lu, and Z.-L. Wang, JHEP, \textbf{1202}, 083 (2012), arXiv:1111.6602[hep-th].

\bibitem{CaramesGuimaraesSilva2012}
T. Carames, M. Guimaraes, and J. H. da Silva, (2012), arXiv:1205.4980[gr-qc].

\bibitem{BazeiaMenezesPetrovSilva2013}
D. Bazeia, R. Menezes, A. Y. Petrov, and A. da Silva, Phys. Lett., B \textbf{726}, 523 (2013), arXiv:1306.1847[hep-th].

\bibitem{BazeiaLobaoMenezesPetrovSilva2014}
D. Bazeia, J. Lobo, A.S., R. Menezes, A. Y. Petrov, and A. da Silva, Phys. Lett., B \textbf{729}, 127 (2014), arXiv:1311.6294[hep-th].

\bibitem{DeWolfeFreedmanGubserKarch2000}
O. DeWolfe, D. Z. Freedman, S. S. Gubser, and A. Karch, Phys. Rev. D, \textbf{62}, 046008 (2000), arXiv:hep-th/9909134.

\bibitem{AfonsoBazeiaLosano2006}
V. I. Afonso, D. Bazeia, and L. Losano, Phys. Lett. B, \textbf{634}, 526 (2006), arXiv:hep-th/0601069.

\bibitem{LiuYangZhaoFuDuan2009}
Y.~X. Liu, J. Yang, Z.~H. Zhao, C.~E. Fu, and Y.~S. Duan, Phys. Rev. D, \textbf{80}, 065019 (2009), arXiv:0904.1785[hep-th].


\bibitem{Lykken1999nb}
  J.~D.~Lykken and L.~Randall,
  JHEP {\bf 0006}, 014 (2000),
arXiv:hep-th/9908076.

\bibitem{Dvali:2000rv}
  G.~R.~Dvali, G.~Gabadadze, and M.~Porrati,
  Phys.\ Lett.\ B {\bf 484}, 112 (2000),
arXiv:hep-th/0002190.

\bibitem{Csaki:2000fc}
  C.~Csaki, J.~Erlich, T.~J.~Hollowood, and Y.~Shirman,
  Nucl.\ Phys.\ B {\bf 581}, 309 (2000),
arXiv:hep-th/0001033.

\bibitem{Csaki:2000pp}
  C.~Csaki, J.~Erlich, and T.~J.~Hollowood,
  Phys.\ Rev.\ Lett.\  {\bf 84}, 5932 (2000),
arXiv:hep-th/0002161.

\bibitem{Long:2002wn}
  J.~C.~Long, H.~W.~Chan, A.~B.~Churnside, E.~A.~Gulbis, M.~C.~M.~Varney, and J.~C.~Price,
  Nature {\bf421}, 922 (2003),
  arXiv:hep-ph/0210004.
\end{thebibliography}
%

\end{document}